\documentclass[a4paper]{PoS}
\usepackage{natbib}
\usepackage{xspace}
\usepackage{amsmath}

\newcommand{\Fermi}{\emph{Fermi}\xspace}

\newcommand{\hsi}{\emph{RHESSI}\xspace}
\newcommand{\goes}{\emph{GOES}\xspace}

\newcommand{\stereo}{\emph{STEREO}\xspace}
\newcommand{\sdo}{\emph{SDO}\xspace}
\def\Rs{$R_\odot$\xspace}
\def\de{$^{\circ}$\xspace}
\def\Angst{$\buildrel _{\circ} \over {\mathrm{A}}$}

\newcommand{\apj}{Astrophys. J.}
\newcommand{\apjs}{Astrophys. J. Supp.~Series}
\newcommand{\apjl}{Astrophys. J. Lett.}

\newcommand{\solphys}{Sol. Phys.}
\newcommand{\aap}{Astronomy and Astrophysics}

\title{Fermi Large Area Telescope observations of high-energy gamma-ray emission from behind-the-limb solar flares}

\ShortTitle{LAT behind-the-limb flares}

\author{\speaker{Melissa Pesce-Rollins}$^{ab}$, Nicola Omodei$^{b}$, Vah\'{e} Petrosian$^{b}$, Wei Liu$^{b}$, Fatima Rubio da Costa$^{b}$ and Alice Allafort$^{b}$ on behalf of the \Fermi-LAT Collaboration\\
  \llap{$^a$}Istituto Nazionale di Fisica Nucleare, Sezione di Pisa, I-56127 Pisa, Italy\\
  \llap{$^b$}W.W. Hansen Experimental Physics Laboratory and Physics Department, Stanford University, Stanford, CA 94305, USA\\
        E-mail: \email{melissa.pesce.rollins@pi.infn.it}\\ \email{nicola.omodei@stanford.edu}\\\email{vahep@stanford.edu}\\ \email{weiliu@lmsal.com} \\\email{frubio@stanford.edu}\\\email{allafort@stanford.edu}}

\abstract{\Fermi-LAT >30 MeV observations have increased the number of detected solar flares by almost a factor of 10 with respect to previous space observations. These sample both the impulsive and long duration phases of \emph{GOES} M and X class flares. Of particular interest is the recent detections of three solar flares whose position behind the limb was confirmed by the \stereo-B spacecraft. While gamma-ray emission up to tens of MeV resulting from proton interactions has been detected before from occulted solar flares, the significance of these particular events lies in the fact that these are the first detections of >100 MeV gamma-ray emission from  footpoint-occulted flares. We will present the \Fermi-LAT, \hsi and \stereo observations of these flares and discuss the various emission scenarios for these sources and implications for the particle acceleration mechanisms.
}

\FullConference{The 34th International Cosmic Ray Conference,\\
		30 July- 6 August, 2015\\
		The Hague, The Netherlands}

\begin{document}

\section{Introduction}
During its first seven years in orbit, the \Fermi Large Area
Telescope~\citep{LATPaper} has detected $>$30 MeV gamma-ray emission
from more than 40 solar flares, nearly 10 times more than
EGRET~\citep{Thompson:93} onboard the {\it Compton Gamma-Ray Observatory},
GRS~\citep{forr85} onboard the {\emph{Solar Maximum Mission} (\emph{SMM})} and
CORONAS-F~\citep{coronasf}. The \Fermi detections sample both the
impulsive~\citep{2012ApJ...745..144A} and the long-duration
phases~\citep{0004-637X-787-1-15} including the longest extended emission ever
detected ($\sim$20 hours) from the SOL2012-03-07 \goes X-class
flares~\citep{0004-637X-789-1-20}. \Fermi-LAT has also provided the first detections of $>$100 MeV emission from three occulted solar flares. These observations sample flares from active regions originating from behind both the eastern and western limbs and include an event associated with the second ground level enhancement event (GLE) of the 24$^{th}$ Solar Cycle. These detections present a unique opportunity to diagnose the mechanisms of high-energy emission and particle acceleration in solar flares.

\section{Behind-the-limb flares detected by the \Fermi-LAT}
In this section we briefly present the observational overview for these events. Table~\ref{tab:BTLQuantities} lists the start times, estimated \emph{GOES} classes, Active Region (AR) locations, CME speeds and total energy released in $>$100 MeV gamma-rays for the three behind-the-limb flares detected by \Fermi-LAT. 
\begin{table}[ht]
  \begin{tabular}{ccccc}
    \hline
    Date & Estimated \emph{GOES} & AR position & CME speed & Total energy \\
    (UTC)  & class  &  & (km s$^{-1}$) & (ergs)\\
    \hline
  2013 Oct 11 07:01 & M4.9  & N21E106 (10$^{\circ}$ b.t.l) & 1200 & 3$\times$10$^{23}$\\
  2014 Jan 06 07:40  & X3.5  & S8W110 (20$^{\circ}$ b.t.l) &  1400 & 5$\times$10$^{21}$\\
  2014 Sep 01 11:00 & X2.1  & N14E126 (36$^{\circ}$ b.t.l) &  2000 & 2$\times$10$^{25}$ \\
  \hline \hline\\
\end{tabular}
\caption{Behind-the-limb flare properties. The start time of the flare is estimated from \stereo imaging. The \emph{GOES} class is evaluated using \stereo-observed extreme ultraviolet intensity, as described by~\citep{2013SoPh..288..241N}. The CME speed listed here is the 2nd-order speed at 20 \Rs taken from the LASCO online catalog and the total energy released is for the >100 MeV LAT emission assuming a smooth time profile.}
\label{tab:BTLQuantities}
\end{table}

\subsection{SOL2013-10-11}
\begin{figure}[ht]
  \begin{center}
    \includegraphics[width=0.49\textwidth,height=0.49\textheight]{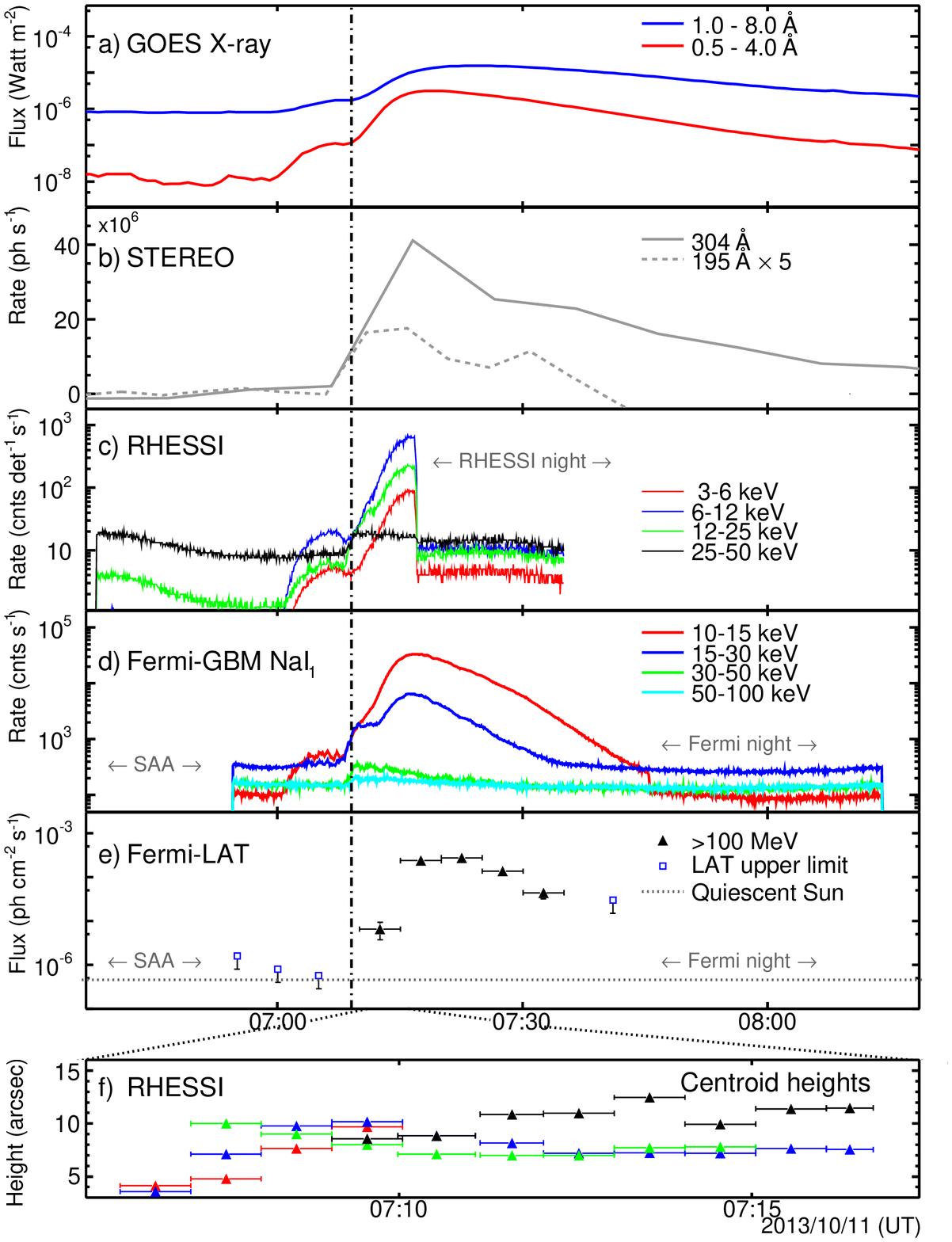}
  \includegraphics[width=0.48\textwidth,height=0.49\textheight]{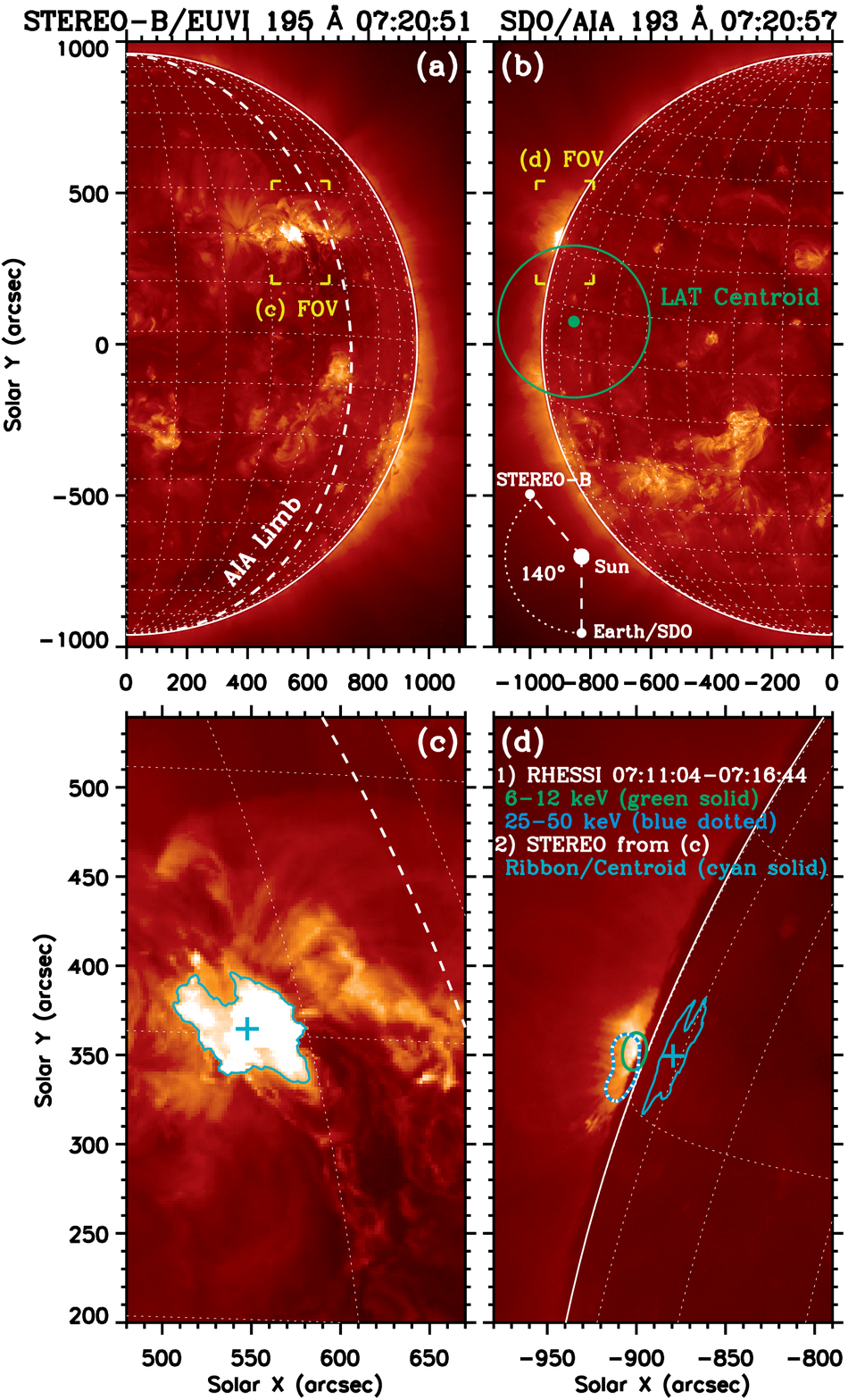}
\caption{Left panel: Light curves of the 2013 October 11 flare~\citep{2041-8205-805-2-L15} as detected by a)
\emph{GOES}, b) \stereo, c) \hsi, d) GBM, e) LAT, and heights of the \hsi
emission centroid (f) with the same color coding as in c). \Fermi exited the
South Atlantic Anomaly (SAA) at 06:57:00 UT. The vertical dashed line represents
flare start time (7:01 UT). Right panel: \stereo-B (left) and \sdo (right) images near the flare peak. The
white-dashed line in (a) and (c) represents the solar limb as seen by \sdo. The
green line in (b) shows the 68\% error circle for the LAT emission centroid. The
cyan contour and plus sign in (c) mark the {\it STEREO} flare ribbon and its
centroid, respectively. Their projected view as seen from the AIA perspective is
shown in (d), in which the centroid is located at 10\de behind the limb. The
green and blue-dotted contours in (d) show \hsi sources. The rectangular
brackets in (a) and (b) mark the field of view (FOV) for (c) and (d),
respectively.}
\label{fig:SOL20131011}
\end{center}
\end{figure}

On 2013 October 11 at 07:01 UT a \emph{GOES} M1.5 class flare occurred with soft X-ray
emission lasting 44 min and peaking at 07:25:00 UT. The left panel of Figure~\ref{fig:SOL20131011} shows the \emph{GOES}, \stereo-B, \hsi, \Fermi Gamma-ray Burst
Monitor~\citep[GBM;][]{meeg09} and LAT lightcurves of this flare. The LAT
detected $>$100~MeV emission for $\sim$30 min with the maximum of the flux
occurring between 07:20:00--07:25:00 UT. \hsi coverage was from
07:08:00--07:16:40 UT, overlapping with \Fermi for 9 min. Based on the \stereo-B 195 \Angst\ we estimate the GOES class of this flare if the active region had not been occulted to be an M4.9.

Images in the right panel of Figure~\ref{fig:SOL20131011} from the \stereo-B Extreme UltraViolet Imager~\citep[EUVI;][]{EUVIinstrument} and the \sdo Atmospheric Imaging Assembly~\citep[AIA;][]{AIApaper} of the photosphere indicate that the active region (AR) was $\sim$10\de behind the limb at the time of the flare. From these images  we find that the off-limb \hsi source lies within the 68\% error circle of the LAT emission. This, however, does not exclude that the emission originates from the visible side of the solar disk.

\subsection{SOL2014-01-06}
\begin{figure}[ht]
  \begin{center}
    \includegraphics[width=0.96\textwidth]{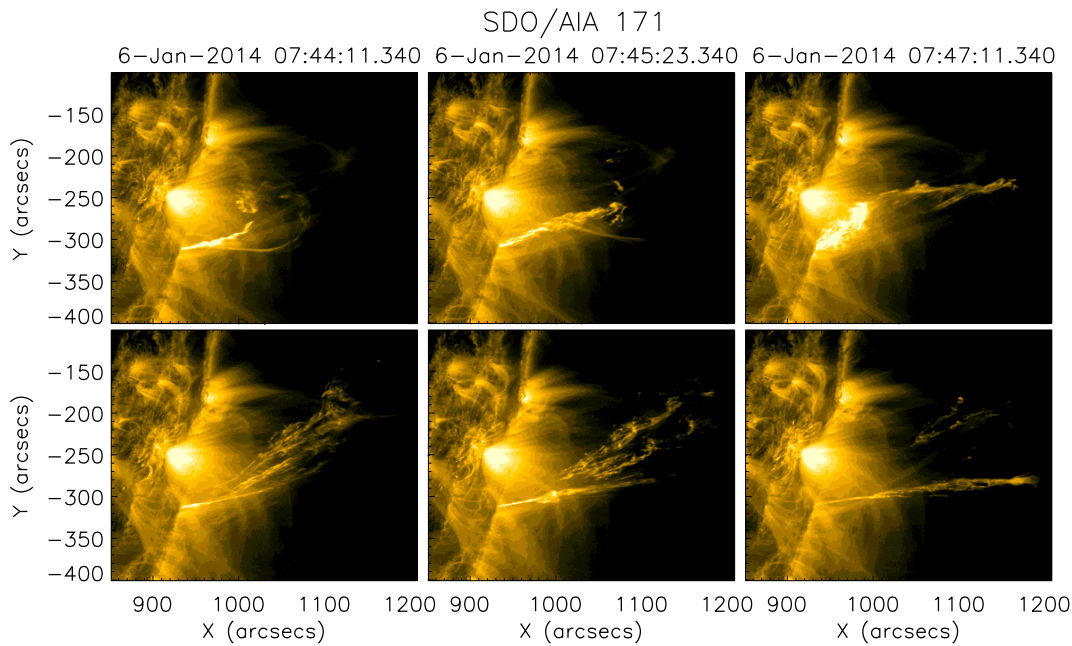}
 \caption{Evolution of the filament eruption associated to SOL2014-01-06 as seen from \sdo/AIA 171.}
\label{fig:SOL20140106}
\end{center}
\end{figure}

On 2014 January 6 a solar flare occurred at approximately 07:42:00 UT from the AR located S8W110 ($\sim$20$^{\circ}$ behind-the-limb). Both \stereo spacecraft had a full view of the AR and detected a large filament erupting from the AR starting from 07:44:00 UT. Figure~\ref{fig:SOL20140106} shows the evolution of this filament as seen from the visible solar disk by \sdo/AIA 171. LASCO detected a halo CME with first C2 appearance at 08:00:00 UT with a 2nd-order Speed at 20 \Rs of 1385 km/s. 
A ground level enhancement (GLE) was also reported in association with this flare making it the second GLE of Solar Cycle 24.  \citet{2014ApJ...790L..13T} report that this event (GLE72) was primarily observed by the South Pole neutron monitors with an increase of $\sim$2.5\% over the background level. The increase in the \emph{GOES} proton flux was evident up to energies $>$700 MeV. Based on the onset times of the SEPs detected at Earth combined with the onset time of the GLE it is possible to estimate the particle acceleration time at the Sun. We find that a straight line fit to the onset times as a function 1/$\beta$ gives the acceleration time to be on 2014 January 6 at 07:49 $\pm$ 3 minutes UT. This is in agreement with the solar particle release time of 07:47 UT reported by \citet{2014ApJ...790L..13T}.

\stereo-B HET data show no significant increase in the proton fluxes during this event indicating that the magnetic connection between the CME and Earth was privileged with respect to that of the \stereo spacecraft. The peak rate detected by \stereo was 2.5$\times$10$^{5}$ photons s$^{-1}$ in its 195 \Angst\ channel, corresponding to a \emph{GOES} X3.5 if it had not been occulted. 

Upon exiting the the SAA at  07:55:00 UT both detectors onboard \Fermi detected emission from this flare. The LAT detected $>$100 MeV emission for approximately  20 minutes and the GBM dected emission in the 10's of keV range. No evidence of extended $>$100 MeV emission was detected by the LAT. \hsi detected off-limb emission starting around 8:20 UT (upon exiting the SAA) also in the 10's of keV for over 40 minutes from this flare.

\subsection{SOL2014-09-01}
On 2014 September 1 a bright solar flare occurred from AR 12158 which was located at N14E126 ($\sim$36$^{\circ}$ behind the limb).  LASCO detected a halo CME with first C2 appearance at 11:12:00 UT with a 2nd-order Speed at 20 \Rs of 1216 km/s. A Type II radio burst with an estimated velocity of 2079 km/s was also measured in association with this flare. \stereo-B had an unblocked view of the entire flare and detected a maximum rate of 1.7$\times$10$^{7}$ photons s$^{-1}$ in its 195 \Angst\ channel, corresponding to a \emph{GOES} X2.1 class if it had not been occulted~\citep{2013SoPh..288..241N}. 
\begin{figure}[ht]
  \begin{center}
    \includegraphics[trim=1cm 2cm 2cm 2cm, clip=true, width=0.49\textwidth]{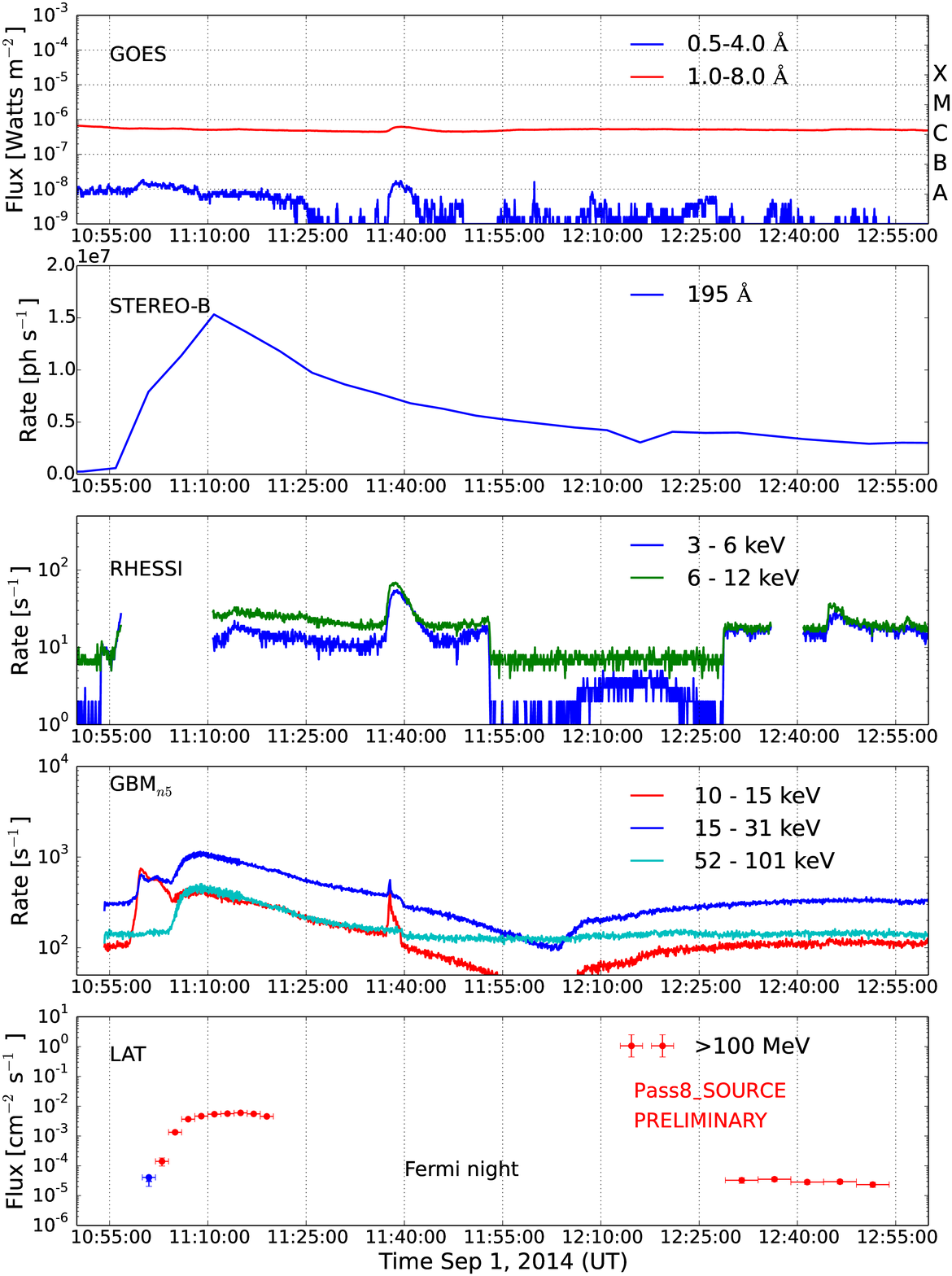}
    \includegraphics[height=0.28\textheight, width=0.49\textwidth]{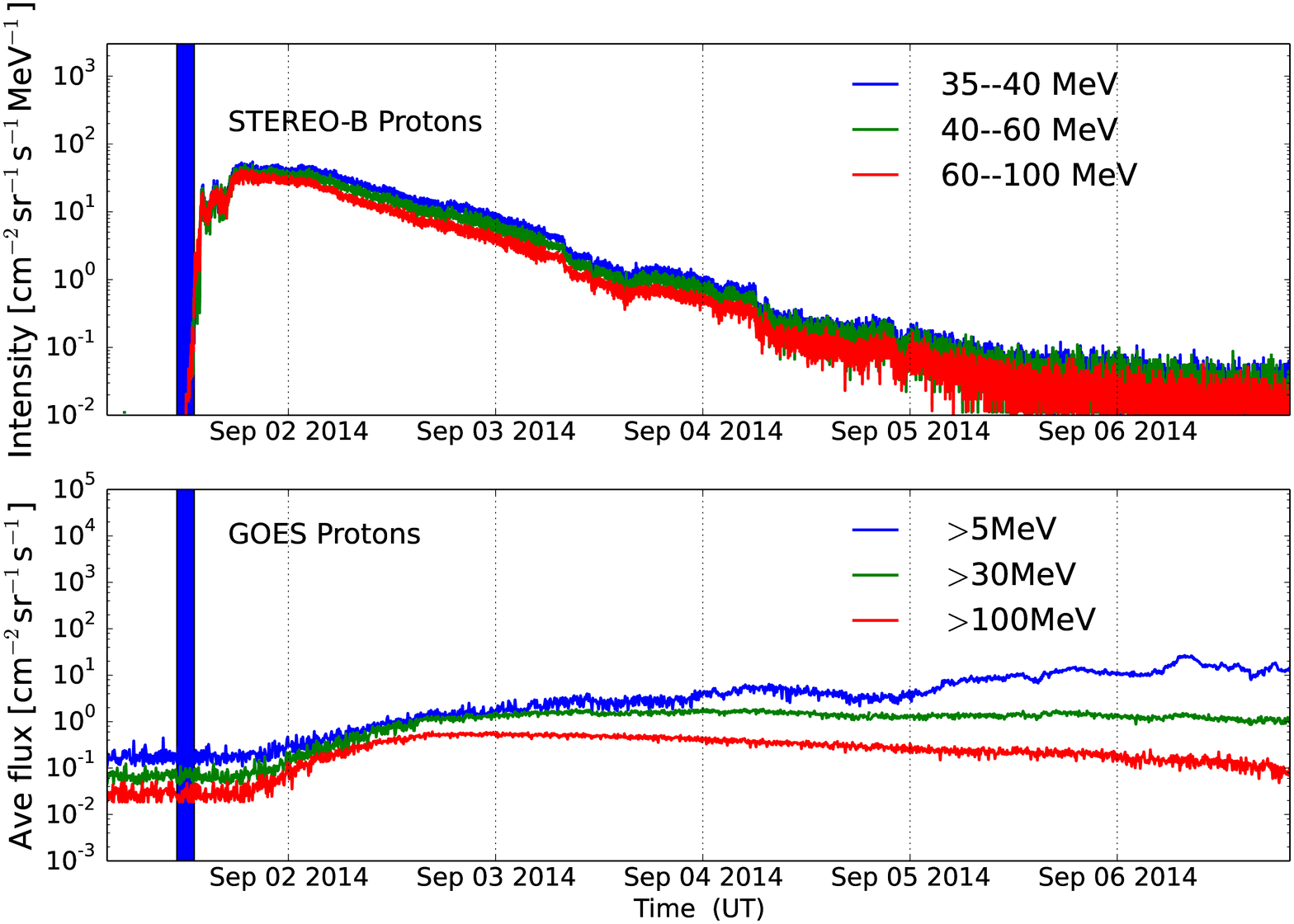}
 \caption{Left panel: Composite lightcurve of \stereo-B EUV, \emph{GOES} X-ray, \hsi, \Fermi-GBM and \Fermi-LAT data for SOL2014-09-01. Right panel: \stereo-B HET proton and \emph{GOES} SEP proton flux.  The blue band in the right hand panel represents the time interval in which the \Fermi-LAT detected gamma-ray emission.}
\label{fig:SOL20140901}
\end{center}
\end{figure}
The composite lightcurve of \stereo-B, \emph{GOES} X-ray, \hsi, \Fermi-GBM and \Fermi-LAT data for this flare is shown in Figure~\ref{fig:SOL20140901}. The high-energy flare emission lasts for $\sim$2 hours but the LAT detected $>$100~MeV emission for $\sim$1 hour due to the \Fermi night between 11:30 and 12:25 UT. The peak LAT flux occurred between 11:10:00--11:15:00 UT. Thanks to improvements provided by the new LAT event selection (\texttt{Pass8}\footnote{A summary of the \Fermi-LAT \texttt{Pass8} performance can be found here \emph{http://www.slac.stanford.edu/exp/glast/groups/canda/lat\_Performance.htm}.}) we gained 10 minutes of detected coverage with respect to \texttt{Pass7\_REP} of this flare during the impulsive phase. As shown in the left panel of Figure~\ref{fig:SOL20140901} the \stereo-B 60--100 MeV proton flux starts to increase roughly 3 hours after the start of the flare whereas the \emph{GOES} $>$100 MeV proton flux shows an increase roughly 9 hours later.

The GBM detected emission in temporal coincidence with the LAT emission in both the BGO and NaI instruments. \hsi was in the SAA from 10:55:00 to 11:11:00, upon exiting the SAA it detected emission up to 12 keV.  Imaging in this time interval with \hsi shows a source located at approximately heliocentric coordinates [$-$900$''$,400$''$] which is in the vicinity of the LAT time integrated $>$100 MeV emission centroid. If the \hsi source is the looptop of the behind-the-limb flare then the minimum height needed for this source to be visible from $\sim$40$^{\circ} $ behind-the-limb would be $\sim$10$^{10}$cm. In fact, X-ray emission from a flare located $\sim$40$^{\circ} $ behind-the-limb has been detected before by \hsi~\citep{KruckerS_Coronal60keV_2007ApJ...669L..49K}.

The LAT measured 18 \texttt{P8R2\_SOURCE} photons with energies $>$1 GeV and reconstructed direction less than 1$^{\circ}$ from the center of the solar disk. 15 of these, including a 3.5 GeV photon, arrived between 11:08 and 11:20 UT.

\section{Discussion}
We present the \Fermi-LAT observations of high-energy gamma-ray emission from three behind-the-limb solar flares (SOL2013-10-11, SOL2014-01-06 and SOL2014-09-01) together with multiwavelength data from \stereo, \emph{GOES} and \hsi.  Based on our work in~\citet{2041-8205-805-2-L15} we can exclude the scenario in which the gamma-ray emission occurred at the AR position due to the large optical depths involved.
Although the scenario in which the emission originates from the Corona is not entirely excluded, we suggest that the gamma-ray emission detected by the LAT was caused by protons accelerated at the CME-driven shock. In this scenario, also proposed by~\citet{1993ICRC....3...91C}, the particles are accelerated on open field lines and can either precipitate to the visible disk and produce gamma-ray emission or escape to be observed as SEPs. Further studies on the precipitation rate of the protons producing the observed gamma-ray emission from these behind-the-limb flares is currently underway together with detailed comparison of the flare characteristics with other on-disk \Fermi-LAT detections (paper in preparation).

\acknowledgments
The \Fermi LAT Collaboration acknowledges support from a number of agencies and
institutes for both development and the operation of the LAT as well as
scientific data analysis. These include NASA and DOE in the United States,
CEA/Irfu and IN2P3/CNRS in France, ASI and INFN in Italy, MEXT, KEK, and JAXA in
Japan, and the K.~A.~Wallenberg Foundation, the Swedish Research Council and the
National Space Board in Sweden. Additional support from INAF in Italy and CNES
in France for science analysis during the operations phase is also gratefully
acknowledged. V.P, W.L and F.R.d.C are supported by NASA grants NNX14AG03G, NNX13AF79G and NNX12AO70G. M.P.R is supported by NASA grant NNX13AC47G.

\end{document}